%Paper: hep-ph/9209204
%From: savage@thepub.ucsd.edu (martin john savage)
%Date: Tue, 1 Sep 92 14:53:54 PDT

\input harvmac
%%%%%%%%%%%%%%%%%%%%%%%%%%%%%%%%%%%%%%%%
%%%%%%%%%%%%%%%%%%%%%%%%%%%%%
%
%  UCSD macros to overwrite some of the definitions in harvmac.tex
%  (include after harvmac.tex)
%  last modified 4/92
%
%%%%%%%%%%%%%%%%%%%%%%%%%%%%%%%%%%%%%%%%
%%%%%%%%%%%%%%%%%%%%%%%%%%%%%%%
%
% modify the output routine for the little format
%
\ifx\answ\bigans
\else
\output={
  \almostshipout{\leftline{\vbox{\pagebody\makefootline}}}\advancepageno
}
\fi
%
%
% address
%
\def\mayer{\vbox{\sl\centerline{Department of Physics 0319}%
\centerline{University of California, San Diego}
\centerline{9500 Gilman Drive}
\centerline{La Jolla, CA 92093-0319}}}
%
% grant numbers
%

%
% preprint number
%
% restores pagenumbers
%
% abstract
%
\def\abstract#1{\centerline{\bf Abstract}\nobreak\medskip\nobreak\par #1}
%
%
% titlefont
%
%
\edef\tfontsize{ scaled\magstep3}
 \tfontsize  \tfontsize
 \tfontsize \font\titlei=cmmi10 \tfontsize
\font\titleis=cmmi7 \tfontsize \font\titleiss=cmmi5 \tfontsize
\font\titlesy=cmsy10 \tfontsize \font\titlesys=cmsy7 \tfontsize
\font\titlesyss=cmsy5 \tfontsize  \tfontsize
\skewchar\titlei='177 \skewchar\titleis='177 \skewchar\titleiss='177
\skewchar\titlesy='60 \skewchar\titlesys='60 \skewchar\titlesyss='60
%
%\def\titlefont{\def\rm{\fam0\titlerm}% switch to title font
%\textfont0=\titlerm \scriptfont0=\titlerms \scriptscriptfont0=\titlermss
%\textfont1=\titlei \scriptfont1=\titleis \scriptscriptfont1=\titleiss
%\textfont2=\titlesy \scriptfont2=\titlesys \scriptscriptfont2=\titlesyss
%\textfont\itfam=\titleit \def\it{\fam\itfam\titleit}\rm}
%
%
% math symbols
%
%---------------------------------------------------------------------
%
\def\inv{^{\raise.15ex\hbox{${\scriptscriptstyle -}$}\kern-.05em 1}}
  %prime
\def\lbar{{\lower.35ex\hbox{$\mathchar'26$}\mkern-10mu\lambda}} %lambda
bar

%
%
% various slashed symbols
%
%
 % slashes a character
\def\dsl{\,\raise.15ex\hbox{/}\mkern-13.5mu D} %this one can be subscripted
\def\delsl{\raise.15ex\hbox{/}\kern-.57em\partial}
\def\Ksl{\hbox{/\kern-.6000em\rm K}}
\def\Asl{\hbox{/\kern-.6500em \rm A}}
\def\Dsl{\hbox{/\kern-.6000em\rm D}} %roman D
\def\Qsl{\hbox{/\kern-.6000em\rm Q}}
\def\gradsl{\hbox{/\kern-.6500em$\nabla$}}
%
% space and backspace in l mode
%
\def\lspace{\ifx\answ\bigans{}\else\qquad\fi}
\def\lbspace{\ifx\answ\bigans{}\else\hskip-.2in\fi} % $$\lbspace...$$
%
%     boxes an equation
%
\def\boxeqn#1{\vcenter{\vbox{\hrule\hbox{\vrule\kern3pt\vbox{\kern3pt
        \hbox{${\displaystyle #1}$}\kern3pt}\kern3pt\vrule}\hrule}}}
%
%     draw a little box (end of proof symbol)
%     e.g. \mbox{.1}{.1}
%
\def\mbox#1#2{\vcenter{\hrule \hbox{\vrule height#2in
\kern#1in \vrule} \hrule}}
%
%
%
%     curly letters
%
   %curly letters
\def\CA{{\cal A}}  \def\CC{{\cal C}} \def\CD{{\cal D}}
   \def\CH{{\cal H}}

%
%
%
%     derivatives
%
%

%

\def\bar#1{\overline{#1}}

\def\darr#1{\raise1.5ex\hbox{$\leftrightarrow$}\mkern-16.5mu #1}

%
 %pound sterling
%
 %puts a small half in a displayed eqn
\def\frac#1#2{{\textstyle{#1\over #2}}} %puts a small fraction
%in a displayed eqn
%
%
%     various math operators
%
%

\def\Tr{\mathop{\rm Tr}}

\def\GeV{{\rm GeV}}

\def\fm{{\rm fm}}
%
%
%
%

%
%       relations
%
\def\ltap{\ \raise.3ex\hbox{$<$\kern-.75em\lower1ex\hbox{$\sim$}}\ }
\def\gtap{\ \raise.3ex\hbox{$>$\kern-.75em\lower1ex\hbox{$\sim$}}\ }
\def\gl{\ \raise.5ex\hbox{$>$}\kern-.8em\lower.5ex\hbox{$<$}\ }
\def\roughly#1{\raise.3ex\hbox{$#1$\kern-.75em\lower1ex\hbox{$\sim$}}}
%
%
%       This defines et al., i.e., e.g., cf., etc.

%

%

\def\pl#1#2#3{{Phys. Lett. } {#1}B (#2) #3}

\relax

\def\dpart{\partial\kern .5ex\llap{\raise
1.7ex\hbox{$\leftrightarrow$}}\kern -.7ex {_\mu}}

\def\frac#1#2{{\textstyle{#1 \over #2}}}

\def\s2weak{\sin^2\theta_{\rm w}}

\def\({\left(}\def\){\right)}

\def\mayer{\vbox{\sl\centerline{Department of Physics}
\centerline{9500 Gilman Drive 0319}
\centerline{University of California, San Diego}
\centerline{La Jolla, CA 92093-0319}}}

\def\Queens{\vbox{\sl\centerline{Department of Physics}
\centerline{Stirling Hall}
\centerline{Queen's University}
\centerline{Kingston, Canada, K7L 3N6}}}

\def\[{\left[}
\def\]{\right]}
\def\({\left(}
\def\){\right)}

\noblackbox
%\draftmode
\vskip 1.in
\centerline{{\titlefont{Electromagnetic Polarisability of the Nucleon}}}
\medskip
\centerline{{\titlefont{in Chiral Perturbation Theory}}}
\bigskip
\vskip .5in
\centerline{Malcolm N. Butler}
\bigskip
\Queens
\bigskip\bigskip
\centerline{Martin J. Savage\footnote{$^{\dagger}$}
{SSC Fellow}}
\bigskip
\mayer
\bigskip\bigskip
\centerline{{\bf Abstract}}
\bigskip

We compute the polarisability of the nucleon to leading order in chiral
perturbation theory.
The contributions from kaons and baryon resonances as intermediate states
are
included
in addition to the contribution from pions and nucleons that had been
previously computed.
The isoscalar operators are dominated by the infrared behaviour of pion loops
giving rise to a $1/m_{\pi}$ coefficient.
In contrast, the  isovector operators are dominated by loops involving kaons,
giving a $1/m_{k}$ coefficient, and further demonstrates that the strange
quark
is an important component of the nucleon.
In addition, the inclusion of the decuplet of baryon resonances as
intermediate
 states substantially modifies the result found from the octet baryons alone
for the isoscalar polarisability.

\vfill
\hbox{\hbox{ UCSD/PTH 92-30\hskip 2in August 1992} }
\hbox{\hbox{ QUSTH-92-04} }
\eject

The electromagnetic polarisability of the nucleon is a unique probe of the
structure of the nucleon.
Like the charge radius and anapole moment,  it is dominated by long-distance
physics in the form of pion and kaon loops and diverges in the chiral limit as
$1/m$ ($m$ being the mass of the pion or kaon).
Recent experimental measurements of the both the proton
\ref\exptp{F.J. Federspiel {\it et al},  Phys. Rev. Lett. {\bf 67}, (1991)
1511.}
and neutron polarisabilities
\ref\exptn{J. Schmiedmyer {\it et al}, Phys. Rev. Lett. {\bf 66}, (1991) 1015.}
have generated increased theoretical effort to understand at a more
fundemental
level what these measurements tell us about the structure of the nucleon.
Several theoretical approaches have been used to estimate the polarisability
of
the nucleon.
These include; dispersion relations
\ref\dispa{M. Damashek and F.J. Gilman,  Phys. Rev. {\bf D1}, (1970) 1319.}
\ref\dispb{A.I. L'Vov, V.A. Petrunkin and J.A. Startsev,  Sov. J. Nucl. Phys.
{\bf 28}, (1979) 651.},
constituent quark models
\ref\quarka{T.E.O. Ericson, in {\it Interaction Studies in Nuclei}, edited by
H. Joachim and B. Ziegler (North Holland , Amsterdam)
(1975).}\nref\quarkb{G.
Dattoli, G. Matone and D. Prosperi, Nuovo. Cim. {\bf 19}, (1977)
601.}\nref\quarkc{T.E.O. Ericson and J. Hufner, Nuc. Phys. {\bf B57}, (1973)
669.}\nref\quarkd{J.L. Friar, Ann. Phys. (NY) {\bf 95} (1975)
170.}-\ref\quarke{V. A. Petrunkin, Fiz. Elem. Chastits At. Yadra, {\bf 12},
(1981) 692.},
bag models
\ref\baga{P. Hecking and G.F. Bertsch,  Phys. Lett. {\bf B99}, (1981)
237.}\nref\bagb{H. Krivine and J. Navarro, Phys. Lett. {\bf B171}, (1986)
331.}\nref\bagc{V. Bernard, B. Hiller and W. Weise, Phys. Lett. {B205}, (1988)
16.}-\ref\bagd{A. Schafer {\it et al}, Phys. Lett. {\bf B143}, (1984) 323.},
pion cloud models
\ref\clouda{R. Weiner and W. Weise, Phys. Lett. {\bf B159}, (1985) 85.},
the Skyrme model
\ref\skyrmea{N.N. Scoccola and W. Weise, Phys. Lett. {\bf B232} (1989)
287.}\nref\skyrmeb{E.M. Nyman, Phys. Lett. {\bf B142}, (1984)
388.}\nref\skyrmec{M. Chemtob, Nucl. Phys. {\bf A473}, (1987)
613.}-\ref\skyrmed{N.N. Scoccola and W. Weise, Nucl. Phys. {\bf A517}
(1990)
495.}
and chiral perturbation theory
\ref\bkm{V. Bernard, N. Kaiser and U. Meissner, Nucl. Phys. {\bf B373} (1992)
346; Phys. Rev. Lett. {\bf 67}, (1991) 1515.}
\ref\bkkm{V. Bernard {\it et al},  Preprint BUTP-92/15, CRN 92-24, (1992)}
(for a review see \ref\holstein{B. Holstein, Preprint UMHEP-375 (1992)}).
In the context of chiral perturbation theory,
previous computations of the polarisability at one loop have included only pion
and nucleon intermediate states
\bkm\bkkm\  which contribute only to the isoscalar polarisation.
However, it was shown by Jenkins and Manohar
\ref\heavybaryon{E. Jenkins and A. Manohar,
\pl{255}{1991}{558};\pl{259}{1991}{353}.}
and Jenkins
\ref\heavybliz{E. Jenkins, Nucl. Phys. {\bf B368} (1992) 190;Nucl. Phys. {\bf
B375} (1992) 561.}
that it is necessary to include both the octet and decuplet of baryons in loop
computations of the baryon axial currents, masses and weak nonleptonic
decays
in order to be consistent with experimental data.
In this letter we compute the electromagnetic polarisability of the nucleon to
leading order in chiral perturbation theory including the lowest lying decuplet
of baryon resonances, octet of pseudo-goldstone bosons and octet of
baryons as
intermediate states.
We will show that the isovector polarisability is determined at leading order
by kaon loops, which is another example of the role the strange quark plays in
the structure of the nucleon.

The strong interaction dynamics of the pseudo-goldstone bosons associated
with
the breaking of $SU(3)_L\otimes SU(3)_R$ chiral symmetry to $SU(3)_V$
flavour
symmetry are described to leading order in a derivative expansion by the
Lagrange density
\eqn\gold{{\cal L}_M = {f_\pi^2\over 8}Tr[ D^\mu\Sigma^\dagger D_\mu\Sigma
] +
.....\ \ ,}
where $f_\pi \approx135$ MeV is the pion decay constant and $\Sigma$ is the
exponential of the pion field
\eqn\sig{\Sigma = \exp(2iM/f_\pi)\ \ \ \ ,}
with $M$ being the octet of pseudo-goldstone bosons
\eqn\octet{M =
\left(\matrix{{1\over\sqrt{6}}\eta+{1\over\sqrt{2}}\pi^0&\pi^+&K^+\cr
\pi^-&{1\over\sqrt{6}}\eta-{1\over\sqrt{2}}\pi^0&\overline{K}^0\cr
K^-&K^0&-{2\over\sqrt{6}}\eta\cr}\right)\ \ \ \ .}
The covariant derivative is with respect to the $U(1)_Q$ of electromagnetism
\eqn\covar{D_\mu = \partial_\mu + i\CA_\mu [Q, \,\,\, ] \ \ \ \ \ ,}
where $\CA$ denotes the electromagnetic field.
Unlike the pseudogoldstone bosons that have a well defined transformation
under
the chiral symmetry
$\Sigma\rightarrow L\Sigma R^\dagger$, the baryon fields do not have a
unique
transformation.
It is convenient to define the field $\xi$ via $\xi^2=\Sigma$
\eqn\squig{\xi = \exp(iM/f_\pi)\ \ \ \ ,}
that transforms under chiral symmetry as
\eqn\xitransform{\xi \rightarrow L \xi U^\dagger = U \xi R^\dagger ,}
where $U$ is defined implicitly by Eq.~\xitransform,
and vector fields with definite parity
\eqn\av{ V^\mu = {1\over2}\left(\xi D^\mu\xi^\dagger+\xi^\dagger
D^\mu\xi\right),\quad
A^\mu = {i\over2}\left(\xi D^\mu\xi^\dagger-\xi^\dagger D^\mu\xi\right)\ \ \ ,}
where $A_\mu$ transforms homogeneously and $V_\mu$ transforms
inhomogeneously
under chiral symmetry.

As recently emphasized \heavybaryon ,
processes where the baryons are nearly on their mass-shell, i.e. processes
involving soft pions, can be consistently described by a chiral lagrangian
involving fields of definite velocity.
With such boosted fields the chiral lagrangian becomes a expansion in the
small
parameters $k/\Lambda_\chi$ where $\Lambda_\chi$ is the chiral symmetry
breaking scale, $k$ is the residual off-shellness of the nucleon (of order
$m_\pi$ )
and $k/M_B$ where $M_B$ is the mass of the baryon.
Both of these expansion parameters are much less than unity  unlike the
expansion in $P/\Lambda_\chi$ (of order unity) where $P$ is the full
momentum
of the baryon, and each order in the derivative expansion is equally important.
This provides a systematic way of power counting in $1/M_B$ and
$1/\Lambda_\chi$ (for a review see
\ref\jmhungary{E. Jenkins and A. Manohar, Proceedings of the workshop
on``Effective Field Theories of the Standard Model'', ed. U. Meissner, World
Scientific (1992)}).
The strong interaction of the lowest lying octet of baryons with the
pseudo-goldstone bosons is described to lowest order in $1/M_B$, lowest
order
in derivatives and to lowest order in the quark mass matrix by the lagrange
density
\eqn\chirallag{\eqalign{{\cal L}_v^8 &= i \Tr \bar B_v \left(v\cdot \CD
\right)B_v
+2D\, \Tr\, \bar B_v\, S_v^\mu\, \{ A_\mu, B_v \}
+2F\, \Tr\, \bar B_v\, S_v^\mu\, [A_\mu, B_v]\cr}\ \ \ \ ,}
where
\eqn\bv{ B_v = \pmatrix { {1\over\sqrt2}\Sigma_v^0 + {1\over\sqrt6}\Lambda_v
&
\Sigma_v^+ & p_v\cr
\Sigma_v^-& -{1\over\sqrt2}\Sigma_v^0 + {1\over\sqrt6}\Lambda_v&n_v\cr
\Xi_v^- &\Xi_v^0 &- {2\over\sqrt6}\Lambda_v \cr }\ \ \ ,}
and where the covariant chiral derivative is $\CD_\mu = \partial_\mu + [
V_\mu
,\ \  ]$.
Only factors of the residual momentum are generated by derivatives acting on
the baryon field.
The strong interaction coupling constants $F$ and $D$ have been determined
at
one-loop in chiral perturbation theory \heavybaryon\ to be $F = 0.40\pm 0.03$
and $D = 0.61\pm 0.04$.
These values correspond to an axial coupling of $g_A = 1.01\pm 0.06$, much
less
than $1.25$ extracted at tree-level.  However, as we are computing one-loop
graphs,  we use the value of coupling constants extracted at one-loop level
for consistency.

The strong interaction of the lowest lying decuplet of baryon resonances is
described at leading order in  $1/M_B$,  lowest order in derivatives and to
lowest order in the quark mass matrix by the lagrange density
\eqn\ltv{\eqalign{{\cal L}^{10}_v &= - i \,\bar T_v^{\mu} (v \cdot \CD) \
T_{v\,\mu}
+ \Delta m \,\bar T_v^{\mu}\, T_{v\,\mu}
+ \CC\,\left(\bar T_v^{\mu} A_{\mu} B_v + \,\bar B_v A_{\mu}
T_v^{\mu}\right)\cr
&+2\CH \,\bar T_v^{\mu} S_{v \,\nu} A^{\nu}  T_{v \, \mu}}\ \ \ \ ,}
where the elements of $T_v$ are (we will supress the lorentz index and
velocity
subscript)
\eqn\decmem{\eqalign{T^{111} & = \Delta^{++}, \ \ T^{112} =
{1\over\sqrt{3}}\Delta^{+}, \ \
T^{122} = {1\over\sqrt{3}}\Delta^{0}, \ \ T^{222} = \Delta^{-}, \ \ T^{113} =
{1\over \sqrt{3}}\Sigma^{*+}\cr
T^{123} & = {1\over\sqrt{6}}\Sigma^{*0}, \ \ T^{223} =
{1\over\sqrt{3}}\Sigma^{*-}, \ \
T^{133} = {1\over\sqrt{3}}\Xi^{*0}, \ \ T^{233} = {1\over\sqrt{3}}\Xi^{*-}, \ \
T^{333} = \Omega^-}}
and $\Delta m $ is the mass splitting between the decuplet of resonances and
the octet of baryons.
The coupling constant $\CH$ and $\CC$ have been determined \heavybaryon\
to be
$\CH = -1.91\pm 0.7$ and $\CC = 1.53\pm 0.3$.

The electromagnetic polarisability of the nucleon is the analogue of the Stark
and Zeeman effects
of atomic physics.
For long-wavelength photons we can match onto an effective field theory
where
there are four terms at dimension seven in the chiral lagrangian (both
isoscalar and isovector) that can contribute to the electromagnetic
polarisability of the nucleon,
\eqn\polar{\eqalign{{\cal L}_v = &{A\over 2}\,\overline{N}_vN_v
F^{\mu\nu}F_{\mu\nu}
 +{B\over 2}\,\overline{N}_vN_v F^{\mu\alpha}F_\mu^\beta v_\alpha
v_\beta\cr
 + & {C\over 2}\,\overline{N}_v\tau^3N_v F^{\mu\nu}F_{\mu\nu}
 +{D\over 2}\,\overline{N}_v\tau^3N_v F^{\mu\alpha}F_\mu^\beta v_\alpha
v_\beta\ \ \ \ },}
where $A, B, C, D$ are coefficients that are to be determined.
$F^{\mu\nu}$ is the electromagnetic field strength tensor, $N$ is the nucleon
doublet (we only need represent  the operators in $SU(2)_L\otimes SU(2)_R$
for
our purposes)
\eqn\nucleon{N = \left(\matrix{p\cr n}\right)\ \ \ \ ,}
and $\tau^3\rightarrow U\tau^3 U^\dagger$ under chiral symmetry.

Naive power counting allows us to estimate the tree level values of the
coefficients $A$ through $D$.
As they are all dimension seven operators we expect
\eqn\dimana{A\sim B\sim C\sim D\sim {e^2\over 2\Lambda_\chi^3}\sim 0.03
\GeV^{-3}\ \ \ \ .}
It was pointed out in \clouda\ that the pion cloud provides the dominant
contribution to the polarisability and,
as recently discussed in refs.\bkm\bkkm\  , the one-loop pion contribution to
the coefficients is infrared divergent, cut off by $m_\pi$. Therefore the
one-loop contribution to the polarisability is enhanced by
$\sim\pi\Lambda_\chi/m_\pi$ over the tree-level values.
We are now in a position to compute the coefficients $A,B,C,D$ to leading
order
in chiral perturbation theory from the graphs shown in \fig\graphs{Graphs that
contribute to the polarisability of the nucleon at leading order in chiral
perturbation theory.   We have chosen to work in  $v\cdot\epsilon = 0$ gauge
in
which the photon does not couple directly to the baryon at leading order in
$1/M_B$.  A heavy black line denotes a baryon of fixed velocity $v$,  a thin
gray line denotes a meson, and a wiggly line  denotes a photon.}.  Essentially
we are integrating the pion out of the theory, leaving an effective theory of
baryons and photons.

Throughout our computation we will assume that the octet of baryons are
degenerate,  and the decuplet of baryon resonances are degenerate but split
from the octet by $\Delta m$.   However, we explicitly retain the mass of the
pion and kaon.  For simplicity we write $A=A^8+A^{10}$ where $A^8$ is the
contribution from the octet baryons and $A^{10}$ is the contribution from the
decuplet of resonances and similarly for $B,C,D$.
To leading order, the coefficients are
\eqn\coeff{\eqalign{A^8 = & {\pi (D+F)^2\over 12f_\pi^2m_\pi}{e^2\over
16\pi^2}
\left[1+{m_\pi\over 2m_K}\left( {{5\over 3}D^2+3F^2-2DF\over
(D+F)^2}\right)\right]\cr
B^8 = & -{11\over 6}{\pi (D+F)^2\over f_\pi^2m_\pi}{e^2\over 16\pi^2}
\left[1+{m_\pi\over 2m_K}\left( {{5\over 3}D^2+3F^2-2DF\over
(D+F)^2}\right)\right]\cr
C^8 = & {\pi \over 24 f_\pi^2m_K}{e^2\over 16\pi^2}
\left[ -{1\over 3}D^2+F^2+2DF\right]\cr
D^8 = & -{11\over 12}{\pi \over  f_\pi^2m_K}{e^2\over 16\pi^2}
\left[ -{1\over 3}D^2+F^2+2DF\right]\cr}\ \ \ \ ,}
which numerically are
\eqn\cnum{\eqalign{ A^8 = & 6.5\times 10^{-2} \, \GeV^{-3}\cr
B^8 = &- 1.4  \, \GeV^{-3}\cr
C^8 = & 4.4 \times 10^{-3} \, \GeV^{-3}\cr
D^8 = & -9.7 \times 10^{-2} \, \GeV^{-3}}\ \ \ \ \ \ .}
The isoscalar amplitude arising from the octet baryons is $\sim 50$ times
larger than the naive size of the contact term.
The contribution to the isovector amplitude is $\sim 4$ times larger than the
naive size of the contact term.  Formally this is the dominant contribution but
in reality the  incalculable contact term may be significant.
However, it is interesting to note that the isovector amplitude is dominated by
kaon loops and hence strange quark configurations of the nucleon.

At this point it is worth commenting on the results obtained in \bkm .
In their computation pion loops were computed in the fully relativistic theory
that does not have a consistent power counting scheme.  They found an
isovector
contribution proportional to
$\log (m_\pi/M_N)$ which is incorrect as the nucleon mass should never
appear
as an argument of the logarithm.   One might imagine infrared terms such as
$1/M_B\log (m_\pi/\Lambda_\chi)$ generated through loop graphs at higher
order
in the $1/M_B$ expansion.  However, to be consistent one needs to also
include
higher order terms in the $1/\Lambda_\chi$ expansion such as
$1/\Lambda_\chi
\log(m_\pi/\Lambda_\chi)$ that may be generated  through loops.  The only
part
of the computation that is  unambiguous and legitimate to retain are the
leading infrared $\pi/m_\pi$ terms as determined in \bkkm.  Numerically we
would expect these subleading  nonanalytic terms to be a few times larger
than
naive estimates of the contact terms and thus negligible.

The contribution from the decuplet of baryon resonances is simply computed
in
the heavy baryon limit.
As the formulae for the different operators are now functions of both the
meson
mass and baryon mass splitting, we introduce two functions
\eqn\bdefine{\eqalign{
Q(x,y) & = -{2y\over y^2-x^2} + {{2\over 9}y^2 -
{11\over 9}x^2\over (y^2-x^2)^{3\over 2}}\log (\gamma (x,y))\cr
R(x,y) & = {1\over 9}{1\over \sqrt{y^2-x^2}}\log(\gamma (x,y))\cr
\gamma (x,y) & = {y-\sqrt{y^2-x^2+i\epsilon}\over y+\sqrt{y^2-x^2+i\epsilon}}}\
\ \ \ .}
We find that the contribution to the coefficients $A,B,C,D$ are
\eqn\codec{\eqalign{
A^{10}  & = -{1\over 2}{\CC^2\over f_\pi^2}{e^2\over 16\pi^2}
\left[ {4\over 3} R(m_\pi, \Delta m) + {1\over 4}R(m_K,\Delta m)\right]\cr
B^{10} & = {\CC^2\over f_\pi^2}{e^2\over 16\pi^2}
\left[ {4\over 3} Q(m_\pi,\Delta m) + {1\over 4}Q(m_K,\Delta m)\right]\cr
C^{10}  & = {1\over 24}{\CC^2\over f_\pi^2}{e^2\over 16\pi^2}
R(m_K,\Delta m)\cr
D^{10} & = -{1\over 12}{\CC^2\over f_\pi^2}{e^2\over 16\pi^2}
Q(m_K,\Delta m)\cr}\ \ \ \ .}
It is not sufficient to keep the leading term in a power series expansion in
$\Delta m/m_\pi$ or $\Delta m/m_K$ as
$\Delta m$, $m_\pi$ and $m_K$ are all comparable in magnitude and so we
retain
the complete functional form.
Numerically we find that
\eqn\decnum{\eqalign{ A^{10} & = 6.4\times 10^{-2} \, \GeV^{-3}\cr
 B^{10} & = -8.9\times 10^{-1}\, \GeV^{-3}\cr
 C^{10} & = -1.5\times 10^{-3}\, \GeV^{-3}\cr
 D^{10} & = 2.9\times 10^{-2}\, \GeV^{-3}\cr}\ \ \ \ .}
We see that the contribution from the decuplet adds to that from the octet for
both isoscalar amplitudes.  However, the contributions from the decuplet and
octet to the isovector amplitudes have the opposite sign and tend to cancel,
but as the octet amplitude is much greater than that from the decuplet, this
effect is small.
We see that it is necessary to include the effect of the decuplet of baryon
resonances as they have a dramatic effect on the predictions.  This should
come
as no surprise as their inclusion is vital as pointed out in ref.\heavybaryon .

Recall that we treated all baryons in the octet as degenerate and all baryons
in the decuplet as degenerate but split from the octet by $\Delta m$.   We
have
investigated the effects of the SU(3) breaking mass difference between the
$\Sigma^*$ and the $\Delta$ and found that it changes the estimate of the
isoscalar amplitudes $A^{10}$, $B^{10}$ by  $\sim 5\%$ but changes the
isovector amplitudes $C^{10}$, $D^{10}$ by as much as $\sim 20\%$.
However,
as the isovector amplitudes induced by octet baryon intermediate states are
much larger than those from the decuplet baryon intermediate states we  feel
confident about treating the baryon multiplets as we have.
Other SU(3) breaking effects may be important, for instance, the coupling
constants $F$, $D$, and $\CC$ were extracted from data in the SU(3) limit.   If
we were to use the value of $\CC$ from the decay $\Delta \rightarrow N\pi$
(as
opposed to the average of $\CC$ found from all measured decays
$T\rightarrow
BM$ where $T$ is a member of the decuplet, $B$ is a member of the octet
and $M$
is a pseudogoldstone boson), we would find the contribution of the decuplet to
be larger than our above estimates.   However, to be consistent all SU(3)
violation must examined at once, including the extraction of coupling
constants
$F$ and $D$.   Until a complete analysis of SU(3) breaking is performed only
the couplings extracted in the SU(3) limit can be used consistently.

To compare our predictions with the experimentally determined values it is
convenient to define the variables $\alpha^{(0)},\alpha^{(1)},\beta^{(0)}$ and
$\beta^{(1)}$-- the isoscalar and isovector electric and magnetic
susceptibilities.
If we use the central values for $F,D,\CC$ we find that
\eqn\results{\eqalign{
\alpha^{(0)}+\beta^{(0)} = & -{1\over 4\pi}(B^8+B^{10}) = 13.9\times 10^{-4} \,
\fm^3\cr
\beta^{(0)} = & -{1\over 2\pi}(A^8+A^{10}) = -0.16\times 10^{-4} \, \fm^3\cr
\alpha^{(1)}+\beta^{(1)} = & -{1\over 4\pi}(D^8+D^{10}) = 0.4\times 10^{-4} \,
\fm^3\cr
\beta^{(1)} = & -{1\over 2\pi}(C^8+C^{10}) = -0.04\times 10^{-4} \, \fm^3
\ \ \ \ ,}}
which are to be compared with the experimentally determined numbers
\exptp\exptn\
\eqn\exptres{\eqalign{
\alpha^{n} = & \, \alpha^{(0)}-\alpha^{(1)} = 12.3\pm 1.5\pm 2.0\times 10^{-4}
\, \fm^3\cr
\beta^{n} = & \, \beta^{(0)}-\beta^{(1)} = 3.5\mp 1.5\mp 2.0\times 10^{-4} \,
\fm^3\cr
\alpha^{p} = & \, \alpha^{(0)}+\alpha^{(1)} = 10.9\pm 2.2\pm 1.4\times 10^{-4}
\, \fm^3\cr
\beta^{p} = & \, \beta^{(0)}+\beta^{(1)} = 3.3\mp 2.2\mp 1.4\times 10^{-4} \,
\fm^3
\ \ \ \ ,}}
where the superscripts $n$ and $p$ denote neutron and proton respectively.
These numbers
seem to be in good agreement with our theoretical estimates , even though
the
experimental errors are still too large to be discriminating.

In conclusion, we have computed the leading contribution to the polarisability
of the nucleon in chiral perturbation theory.
The leading infrared behaviour of the pion graphs had been found before
\bkm\bkkm\  and contribute  only to the isoscalar polarisability.  To be
consistent at the one-loop level the values for $F$ and $D$ were extracted
from
the one-loop computation \heavybaryon\heavybliz\  and also the decuplet of
baryon resonances was included.
As expected, the decuplet had a large effect on the estimate from chiral
perturbation theory and
further, we have found that the leading contribution to the isovector
polarisability arises from kaon loops and hence strange quark configurations
of
the nucleon.
Our results are in remarkable agreement with the measured polarisabilities
and
also the model independent estimate of $\alpha^p+\beta^p = 14.2\pm
0.03\times
10^{-4}\, \fm^3$ found from dispersion relations.
There will be corrections to our results arising from terms higher order in the
loop expansion, $1/M_B$ expansion, $1/\Lambda_\chi$ expansion and from
insertions of the light quark mass matrix.   We expect that these terms are all
small compared with the nonanalytic amplitudes computed in this work for
both
the isoscalar and isovector amplitudes.

The same computation can be done for the spin-dependent amplitudes.
Again we
suspect that
the isovector component will be  dominated by kaon loops and the baryon
resonances will play a vital role.
Similar graphs make the dominant contribution to
the decay $\Sigma^0\rightarrow\Lambda\gamma\gamma$ which proceeds
predominantly
through kaon loops.
This will be the subject of future work.

After this work was completed two preprints
\ref\cohena{T.D. Cohen and W. Broniowski, Preprint, U. of MD PP \#92-191.}
\ref\cohenb{W. Broniowski and T.D. Cohen, Preprint, U. of MD PP \#92-193.}
were released discussing the role of the decuplet of resonances in the
polarisability of the nucleon.

\bigskip\bigskip

We wish to thank E. Jenkins, A. Manohar and M. Luke for numerous
discussions.
We would also like to thank Barry Holstein for giving us the kick in the pants
needed to finish the work.
MNB acknowledges the support of the Natural Science and Engineering
Research
Council  (NSERC) of Canada.
MJS acknowledges the support of a Superconducting Supercollider National
Fellowship from the Texas National Research Laboratory Commission
under grant  FCFY9219.

\listrefs
\listfigs
\end